# Privacy-Preserving Socialized Recommendation based on Multi-View Clustering in a Cloud Environment

Cheng Guo
*School of Software Technology*
*Dalian University of Technology*
Dalian, China
guocheng@dlut.edu.cn

Jing Jia
*School of Software Technology*
*Dalian University of Technology*
Dalian, China
jiajing1995@163.com

Peng Wang
*School of Journalism and Communication*
*Jilin Normal University*
Changchun, China
63999266@qq.com

Jing Zhang
*School of Software Technology*
*Dalian University of Technology*
Dalian, China
2974007859@qq.com

*Abstract*—Recommendation as a service has improved the quality of our lives and plays a significant role in variant aspects. However, the preference of users may reveal some sensitive information, so that the protection of privacy is required. In this paper, we propose a privacy-preserving, socialized, recommendation protocol that introduces information collected from online social networks to enhance the quality of the recommendation. The proposed scheme can calculate the similarity between users to determine their potential relationships and interests, and it also can protect the users' privacy from leaking to an untrusted third party. The security analysis and experimental results showed that our proposed scheme provides excellent performance and is feasible for real-world applications.

*Keywords—multi-view clustering, homomorphic encryption, socialized recommendation*

## I. INTRODUCTION

With the development of the Internet, multiple data are generated every day. However, the torrential flood of information may cause the problem of information overload and cause users to become disoriented. The emergence of recommendation services can help people access what they are really interested in and filter out the meaningless information, thereby enhancing the users' experiences. Also, a good recommendation protocol can attract users' attention and gain high rates of page views and click-throughs, which can also bring economic benefits to the recommendation service providers. Therefore, determining how to construct a good recommendation scheme and determining how to increase the quality of service have become significant areas of research.

There are two main kinds of recommendation models, i.e., content-based models and Collaborative Filtering (CF) models. Content-based recommendation models focus on the properties of items [1][2]. For example, if there is a user who prefers listening to Jazz, the system will recommend more Jazz music to this user. In other words, the content-based model attempts to determine the latent relationship between different items and recommend similar items to users. CF models analyze the historical interactions of users and make predictions by focusing on measuring the similarity between users or items based on interactions [3][4]. For instance, Alice is estimated to be a user who is similar to Bob, and she gave a 5-star evaluation to the movie *The Avengers*, so the system would recommend *The Avengers* to Bob. In recent years, much research on content-based recommendations and the CF model has shown that they are practical for real-world applications [5][6]. However, they still have some limitations.

The problem of content-based recommendations is that traditional schemes do not make full use of the information associated with items in order to determine the latent relevance between them. Specifically, there are many different ways to portray an item, e.g., by describing it in words, using numeric ratings, and based on users' comments. Traditional content-based recommendation schemes usually have considered one single view and ignored the implicit relationship between other views and the abundant information they provided, which resulted in the wasting the available data.

The shortcoming of CF model is that it is overly-dependent on users' historical interactions. For a system in which there have been sparse historical interactions of users, it is difficult to measure the relevance between users and make predictions based on the limited information, which is the so-called "Cold Start" problem. A CF model that has the Cold Start problem will present low-quality recommendations, so that it is very important to eliminate this obstacle.

In order to make full use of information from different perspectives, we combined multi-view clustering with content-based recommendations to determine the implicit relevance of items. The multi-view clustering protocol takes all the views concerning the item into consideration, and it is helpful in finding the nearest neighbors of each item. As for the Cold Start problem, some researchers have discovered that users' social information is valuable for predicting their preferences and enhancing the quality of recommendation services [7-9]. It is convincing that people's choices and decisions probably are influenced by their close friends. However, the topographic maps and historical interactions in social networks contain extensive sensitive information about users, and this information can be used to deduce the users' characteristics and to identify specific people, so such leakage can result in threats to people's property or even their lives. Therefore, the protection of people's privacy should be taken into consideration when introducing social data to recommendation systems.

In this paper, we propose a Privacy-Preserving Socialized Recommendation (PPSR) scheme that can fully determine the implicit information of items and solve the data sparsity problem by using the online social network. In addition, the topographic maps and historical interactions of the social network are kept secure and separate from the recommendation system, so users are provided high-quality services without the violation of their privacy. Our contributions are summarized as follows:

(1) We applied the multi-view clustering protocol to determine the potential correlation between different items to enhance the quality and accuracy of recommendations and to introduce the online social network as a means of solving the Cold Start problem.

(2) We considered several different aspects of users' social data and constructed a unified similarity function to measure the correlation between users.

(3) Our proposed approach protects the confidentiality of social data, which is not the case for existing socialized recommendation methods.

In the next section, a brief introduction to the corresponding techniques is provided. Our proposed approach is presented in Section III, and its performance and security are evaluated in Section IV. We compared its performance with several baseline methods, and, then, our conclusions are presented in Section V.

## II. PRELIMINARIES

### A. Multi-View Clustering

Non-negative Matrix Factorization (NMF) initially was proposed as a dimensionality reduction technology, and it has been used extensively for many purposes, such as recognizing patterns, retrieving information retrieval, and the extraction of features. It can mine the latent semantic relationship extensively to facilitate predictions and recommendations. We used NMF to resolve the multiple different views of data to address the issue of multi-view clustering and to calculate the nearest neighbors of each item in the database.

### B. Homomorphic Encryption

In order to protect the confidentiality of social data and jointly compute the recommendation score of items for a target user, online social network providers encrypt the similarity degrees using the Paillier cryptosystem [10], which supports homomorphic computation.

The Paillier cryptosystem provides additive homomorphism because:
$$E_{pk}(a+b) \leftarrow E_{pk}(a) \times E_{pk}(b) \bmod N^2 \quad (1)$$

It also provides plaintext multiplication:
$$E_{pk}(a \cdot b) \leftarrow E_{pk}(a)^b \bmod N^2 \quad (2)$$

## III. THE PROPOSED SCHEME

In this section, we introduce the details of our Privacy-Preserving Socialized Recommendation (PPSR) scheme.

### A. Overview of PPSR

Our PPSR scheme utilizes multi-view clustering to make full use of the unstructured data, such as the descriptions of items and the comments of users to determine the implicit relationships between items. For newly-registered users who have no historical interactions, PPSR introduces the data of users from the online social network to measure the similarity between users and provide high-quality recommendations. Sensitive data from online social networks are encrypted so that the privacy of users is well protected.

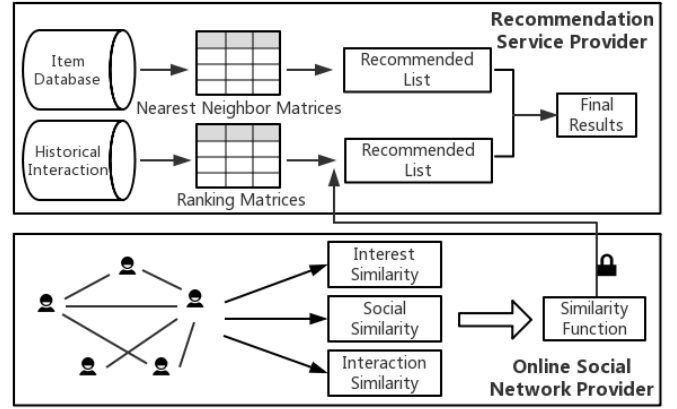

Fig. 1  System model of PPSR

Fig. 1 shows a model of the system and its three phases, which are (1) item recommendation based on multi-view clustering; (2) similarity measurement of users; and (3) privacy-preserving socialized recommendation.

### B. Construction of PPSR

Here, we introduce the details of the construction of our PPSR scheme.

*1) Item recommendation based on multi-view clustering*

Real-world datasets often consist of abundant information from several different domains. For instance, a product on the E-commerce platform has specific categories, content descriptions, customers' comments, and sales volume. These data can be represented as different attributions to depict this product, and each attribution is a single view of the product. It is obvious that data mining based on one single view ignores the information of other views, so making full use of the information provided by multiple different views is worth consideration.

The multi-view clustering method is an extension of the single-view clustering method, and it can mine the implicit semantic information more deeply and measure the similarity between items. In this paper, we use multi-view clustering to explore the hidden relevance and correlation between existing items in the database to improve the accuracy and comprehensibility of the recommendations.

Non-negative Matrix Factorization (NMF) is a feasible approach for learning the latent correlative information of the original data using the following formula:
$$V_{(m \times n)} \approx W_{(m \times K)} \cdot H_{(K \times n)} \quad (3)$$
where $V$, $W$, and $H$ are non-negative matrices, and $K$ is the number of clustering centers defined in advance. Specifically, each row $v_i$ of $V$ is an item of the original database, and $W_{ik}$ denotes the degree of association between item $i$ and the clustering center $k$. Therefore, the goal of NMF-based clustering is to obtain the matrix $W$ to determine which clustering center each item belongs to according to the following formula:

$$O = ||V - WH||, s.t. \ W \geqslant 0, H \geqslant 0 \quad (4)$$

The optimization problem (6) is non-convex because there are two variables $W$ and $H$, and we only can learn one matrix at a time with the other variable fixed:

If we fix $W$, then:

$$H \leftarrow H \frac{W^T V}{W^T W H} \quad (5)$$

If we fix $H$, then:

$$W \leftarrow W \frac{V H^T}{W H H^T} \quad (6)$$

After obtaining the coefficient matrix $W$ for each item $i$, we choose the largest degree of row $i$ and assign item $i$ to the corresponding clustering center.

For the multi-view database, we assume that similar items should be assigned to the same clustering center in the different views. Assume that there are $n_v$ different views of the original database, denoted as $\{V^{(1)}, V^{(2)}, \cdots, V^{(n_v)}\}$, where $V^{(s)} \in \mathbb{R}^{m \times n^{(s)}}$. Each matrix is factorized according to $V^{(s)} \approx W^{(s)} H^{(s)}$, where $W^{(s)} \in \mathbb{R}^{m \times K}$ and $H^{(s)} \in \mathbb{R}^{K \times n^{(s)}}$. In general, the objective function with the constraint $W^{(s)} \geqslant 0, H^{(s)} \geqslant 0$ is shown as:

$$J = \sum_{s=1}^{n_v} \lambda_s ||V^{(s)} - W^{(s)} H^{(s)}|| + R \quad (7)$$

where $\lambda_s$ is a parameter for each view to combine the factorization with other views, and $R(\cdot)$ is the regularization term to improve the constraint performance:

$$R = \sum_{s=1}^{n_v} \sum_{t=1}^{n_v} \lambda_{st} ||W^{(s)} - W^{(t)}||$$

$$= \sum_{s,t} \lambda_{st} ||W^{(s)} - W^{(t)}|| \quad (8)$$

where $\lambda_{st}$ is the weight of the similarity constraint between $W^{(s)}$ and $W^{(t)}$.

If we replace $R$ in formula (7) with formula (8), the final objective function with the constraint $W^{(s)} \geqslant 0, H^{(s)} \geqslant 0$ is:

$$J = \sum_{s=1}^{n_v} \lambda_s ||V^{(s)} - W^{(s)} H^{(s)}||$$
$$+ \sum_{s,t} \lambda_{st} ||W^{(s)} - W^{(t)}|| \quad (9)$$

Similar to single-view clustering, we learn one variable with the other variable fixed. Therefore, the update rules are shown as follows:

$$H^{(s)} \leftarrow H^{(s)} \frac{W^{(s)T} V^{(s)}}{W^{(s)T} W^{(s)} H^{(s)}} \quad (10)$$

$$W^{(s)} \leftarrow W^{(s)} \frac{\lambda_s V^{(s)} H^{(s)} + \sum_{t=1}^{n_v} \lambda_{st} W^{(t)}}{\lambda_s W^{(s)} H^{(s)} H^{(s)T} + \sum_{t=1}^{n_v} \lambda_{st} W^{(s)}} \quad (11)$$

The algorithm is summarized in Algorithm 1.

---

**Algorithm 1** Multi-View Clustering

**Input:**
　　View matrices $\{V^{(1)}, V^{(2)}, \cdots, V^{(n_v)}\}$
　　Parameters $\{\lambda_1, \lambda_2, \cdots, \lambda_{n_v}\}, \lambda_{st}, K$
**Output:**
　　Coefficient matrices $\{W^{(1)}, W^{(2)}, \cdots, W^{(n_v)}\}$
　　Regularize $V^{(s)}$ satisfies $||V^{(s)}|| = 1, s = 1, 2, \cdots, n_v$.
　　Initialize $W^{(s)}, H^{(s)}$ at random, $s = 1, 2, \cdots, n_v$.
　　**repeat**
　　　　Fix $W^{(s)}$, update $H^{(s)}$ by Formula(8),
　　　　$s = 1, 2, \cdots, n_v$;
　　　　Fix $H^{(s)}$, update $W^{(s)}$ by Formula(9),
　　　　$s = 1, 2, \cdots, n_v$;
　　**until** convergency.
　　**return** $\{W^{(1)}, W^{(2)}, \cdots, W^{(n_v)}\}$

---

After applying the multi-view clustering algorithm on the original item database, we obtain the clustering center to which each item belongs, and items in the same class compose the candidate nearest neighbors.

*2) Similarity measurement of users*

For a newly-registered user without historical interactions, it is challenging for the recommendation system to provide a high-quality recommendation service. It is convincing that users of common characteristics are inclined to have similar preferences so that the introduction of an online social network would be helpful in determining the latent relationships among different users and then recommend what they really want. In the online social network, there are several aspects that reflect the properties of users, such as published content, followers, friends, and historical interactions. In this part, we make full use of the information from users' social media and define a similarity function to measure the correlation between users, which is valuable to recommend items for users without historical interactions based on what their nearest neighbors preferred.

In this paper, we focused on three types of information of users from online social networks, i.e., publication information, social connections, and positive interactions. Suppose there are $n_u$ distinct users denoted as $U = \{u_1, u_2, \cdots, u_{n_u}\}$, each of which has a profile consisting of the three types of information mentioned earlier, denoted as $Pr(u_i) = \{P(u_i), C(u_i), I(u_i)\}, u_i \in U$.

$P(u_i)$ represents the publication information of $u_i$, which exists mainly in the form of content and must be transformed into text vectors. The details of the process are as follows: (i) Move away the noise in the content, such as the stop words "a", "an", "the", and some meaningless punctuation marks. (The natural language processing tools NTLK and SCIKIT-LEARN of Python would be helpful.); (ii) Analyze the text and extract keywords from the publications; (iii) Compute the weight of keywords according to the TF-IDF algorithm; (iv) Generate the representation vector $P(u_i) = \{p_{i1}, p_{i2}, \cdots, p_{in_w}\}$, where $p_{ij}$ denotes the weight of the $j^{th}$ keyword in the published content of user $u_i$, and the total number of keywords in the word list is $n_w$.

For two users, i.e., $u_i$ and $u_j$, we use the Cosine Similarity to measure their similarity of interest extracted from what they published:

$$sim(P(u_i), P(u_j)) = \frac{P(u_i) \cdot P(u_j)}{|P(u_i)| \cdot |P(u_j)|}$$

$$= \frac{\sum_{k=1}^{n_w} p_{ik} \times p_{jk}}{\sqrt{\sum_{k=1}^{n_w} p_{ik}^2} \times \sqrt{\sum_{k=1}^{n_w} p_{jk}^2}} \quad (12)$$

$C(u_i)$ represents the social connections of $u_i$. In this paper, we focus on the common friends of different users and who they follow to measure the similarity of their interests because people they follow may reflect what they are interested in, and the number of common friends implies that two users may have similar backgrounds and similar preferences. As mentioned before, there are $n_u$ distinct users, and the relationship between different users can be represented as an $n_u \times n_u$ matrix $R$. Each element of $R$ is set as follows:

$$R_{ij} = \begin{cases} 1 & , if\ u_i\ follows\ u_j \\ 0 & , otherwise \end{cases} \quad (13)$$

For each user $u_i$, the $i^{\text{th}}$ row of $R$ denotes the people who $u_i$ follows. Then, we define another vector $F(u_i) = \{f_{i1}, f_{i2}, \cdots, f_{in_w}\}$ to represent the close friends of $u_i$ (if $R_{ij} = R_{ji} = 1$, we call $u_j$ a close friend of $u_i$, where $f_{ik} = R_{ik} \wedge R_{ki}$.

We calculate the social similarity of two users, i.e., $u_i$ and $u_j$, according to the following formula:

$$sim(C(u_i), C(u_j)) = \lambda_R \frac{R_i \cdot R_j}{|R_i| \cdot |R_j|} + \lambda_F \frac{F(u_i) \cdot F(u_j)}{|F(u_i)| \cdot |F(u_j)|}$$

$$= \lambda_R \frac{\sum_{k=1}^{n_w} R_{ik} \times R_{jk}}{\sqrt{\sum_{k=1}^{n_w} R_{ik}^2} \times \sqrt{\sum_{k=1}^{n_w} R_{jk}^2}}$$

$$+ \lambda_F \frac{\sum_{k=1}^{n_w} f_{ik} \times f_{jk}}{\sqrt{\sum_{k=1}^{n_w} f_{ik}^2} \times \sqrt{\sum_{k=1}^{n_w} f_{jk}^2}} \quad (14)$$

where $\lambda_R$ and $\lambda_F$ are weight parameters.

$I(u_i)$ represents the positive interactions of $u_i$. In this paper, we chose three types of interactions, i.e., like, comment, and repost, denoted respectively as $n_u \times n_u$ matrices $Lk$, $Cmt$, and $Rp$. Specifically, $Lk_{ij}$ ($i \neq j$) denotes the number of published articles for which both $u_i$ and $u_j$ clicked "like" button, and $Lk_{ii}$ denotes the total number of published articles for which $u_i$ clicked "like" button. Similarly, $Cmt$ and $Rp$ also can be generated like $Lk$. The difference is that comments and reposts may contain users' emotion towards the articles so that we analyze the information of comments and reposts and only accumulate the number of positive comments and reposts using the text sentiment analysis algorithm [11].

For two users, $u_i$ and $u_j$, we calculate their interactive similarity according to the following formula:

$$sim(I(u_i), I(u_j))$$
$$= \lambda_{Lk} \frac{Lk_{ij}}{Lk_{ii}} + \lambda_{Cmt} \frac{Cmt_{ij}}{Cmt_{ii}} + \lambda_{Rp} \frac{Rp_{ij}}{Rp_{ii}} \quad (15)$$

where $\lambda_{Lk}$, $\lambda_{Cmt}$ and $\lambda_{Rp}$ are weight parameters.

In general, the final similarity measurement function between two users, $u_i$ and $u_j$, containing their publication information, social connections, and positive interactions is defined as follows:

$$Sim(Pr(u_i), Pr(u_j)) = \lambda_P \cdot sim(P(u_i), P(u_j))$$
$$+ \lambda_C \cdot sim(C(u_i), C(u_j))$$
$$+ \lambda_I \cdot sim(I(u_i), I(u_j)) \quad (16)$$

where $\lambda_P$, $\lambda_C$, and $\lambda_I$ are weight parameters.

*3) Privacy-preserving socialized recommendation*

In the socialized recommendation protocols, the similarity measurement function of users usually is generated by the online social network providers based on social topographic maps and users' interactions. Considering the commercial profits and users' privacy, the original data should be protected from leaking to an untrusted third party. Therefore, the similarity score calculated by the online social network provider (referred to as ***Bob***) must be encrypted before it is transmitted to the recommendation service provider (referred to as ***Alice***). In this paper, we chose Paillier Homomorphic Encryption to protect the similarity scores, and the details are provided below.

Suppose there are $n_u$ distinct users denoted as $U = \{u_1, u_2, \cdots, u_{n_u}\}$, and there are $m$ distinct items denoted as $T = \{t_1, t_2, \cdots, t_m\}$.

First, for a target user $u_i$, Bob computes the similarity scores between $u_i$ and other users, which is $S_{u_i} = \{Sim(Pr(u_i), Pr(u_j)) | u_j \in U \backslash \{u_i\}\}$. Then Bob encrypts $S_{u_i}$ and sends $E_{pk}(S_{u_i})$ to Alice.

As a recommendation service provider, Alice has the historical ranking of items, denoted as the $n_u \times m$ matrix, $Rank$. Each element $Rank_{ij}$ is the ranking score that user $u_i$ gave to item $t_j$. After receiving the similarity scores, Alice can calculate the recommendation degree of each item according to the following formula:

$$E_{pk}(Degree(u_i, t_k))$$
$$= \prod_{u_j \in U \backslash \{u_i\}} E_{pk}(Sim(Pr(u_i), Pr(u_j)))^{Rank_{jk}}$$
$$= \prod_{u_j \in U \backslash \{u_i\}} E_{pk}(Sim(Pr(u_i), Pr(u_j)) \times Rank_{jk})$$
$$= E_{pk}\left(\sum_{u_j \in U \backslash \{u_i\}} \left(Sim(Pr(u_i), Pr(u_j)) \times Rank_{jk}\right)\right) \quad (17)$$

Alice does not have the secret key to decrypt the degrees, so she adds a random number, $r$, to the secure degrees and sends them back to Bob. Bob decrypts the secure degrees and returns a candidate recommendation list named $CRList_{socialized}$ in the ranking from higher degree to lower degree. Each element in $CRList_{socialized}$ is the identification of items, and the decrypted scores are kept secret from Alice. Also, Bob has no idea about which exact item that each id matches.

The process is shown as Algorithm 2.

For a newly-registered user without any historical information, we can execute Algorithm 2 to provide a socialized recommendation service. Once the user shows preferences for some items, we can analyze these items and use Algorithm 1 to determine their nearest neighbors. These neighbors compose a candidate recommendation list, which can be merged with $CRList_{socialized}$ to obtain the final recommendation results.

**Algorithm 2** Privacy-Preserving Socialized Recommendation
**Input:**
    $E_{pk}(S_{u_i})$
    $Rank$
**Output:**
    $CRList_{socialized}$
    **for** $1 \leq k \leq m$ **do**
        Alice Computes:
        $d_k \leftarrow E_{pk}(Degree(u_i, t_k))$
        $d'_k \leftarrow d_k \times E_{pk}(r)$
        $D = D \cup \{d'_k\}$
    **end for**
    Alice sends $D$ to Bob
    Bob decrypts $D$ and ranks the items
    Bob generates $CRList_{socialized}$
    Bob returns $CRList_{socialized}$ to Alice

## IV. THE PROPOSED SCHEME

In this section, we evaluate the performance of our proposed SCMR scheme in terms of security and searching accuracy.

### A. Performance of multi-view clustering

Table I Multi-view clustering results

| Dataset | Algorithm | Acc.(%) | F1(%) | NMI(%) |
|---|---|---|---|---|
| Last.FM | K-Means | 41.1±2.2 | 25.5±1.3 | 41.2±1.3 |
| | SVD | 49.0±0.6 | 38.1±0.5 | 45.6±0.4 |
| | NMF | 46.0±2.5 | 35.8±1.5 | 45.4±1.5 |
| | Multi-C | **50.9±2.7** | **38.7±1.9** | **47.4±1.2** |
| Delicious | K-Means | 40.3±2.1 | 34.4±1.7 | 40.7±1.8 |
| | SVD | 53.1±0.9 | 48.5±1.2 | 48.9±1.6 |
| | NMF | 44.8±4.2 | 42.5±3.3 | 43.6±4.0 |
| | Multi-C | **54.7±3.9** | **49.6±3.1** | **50.2±3.6** |
| Group Leans | K-Means | 52.8±6.9 | 44.5±6.0 | 48.9±6.6 |
| | SVD | 68.4±0.2 | 63.3±0.3 | 64.9±0.1 |
| | NMF | 60.5±5.5 | 58.2±4.4 | 58.2±4.5 |
| | Multi-C | **69.4±6.5** | **64.5±5.3** | **66.3±2.9** |

As described in Section III, multi-view clustering is a practical approach to determine the implicit semantic information in different views of data. In this part, we evaluate multi-view clustering by conducting several experiments using three different datasets:

Last.FM [12]: It is collected from the Delicious social bookmarking system that has 1892 users, 17,632 artists, 92,834 user-listened artist relations, 11,946 tags, and 186,479 tag assignments.

Delicious Bookmarks [12]: It is collected from Last.fm online music system containing 1867 users, 69226 URLs, 104799 bookmarks, 53388 tags and 437593 tag assignments.

GroupLeans [12]: It is an extension of the MovieLens10M dataset, published by the GroupLeans research group. It links the movies of the MovieLens dataset with their corresponding web pages at in the Internet Movie Database (IMDb) and in the Rotten Tomatoes movie review systems. It contains 2,113 users, 10,197 movies, 20 movie genres, 20,809 movie genre assignments, 13,222 tags, 47,957 tag assignments, and 855,598 ratings.

In this part, we chose three classical clustering methods, i.e., K-Means, Singular Value Decomposition (SVD), and single-view NMF. In the experiments, we chose three criteria to evaluate the multi-view clustering performance, i.e., Accuracy, the F1 value, and the Normalized Mutual Information (NMI) value.

Table I lists the Accuracy, the F1 value, and the NMI value for our scheme and three other baseline protocols on three datasets. It is obvious that the clustering ability of our scheme outperformed the other protocols. In other words, information from different views can complement others and completely discover the latent relationships between items.

### B. Performance of recommendation

In order to evaluate the performance of the recommendation performance of the entire Privacy-Preserving Socialized Recommendation (PPSR) model based on the multi-view clustering scheme, we set three models as baseline schemes, i.e., (1) recommendation model based on single-view clustering without social data (called RM-SV); (2) recommendation model based on multi-view clustering without social data (called RM-MV); (3) recommendation model based on single-view clustering with social data (called RM-SVS).

We conducted experiments using two datasets, i.e., Last.FM (mentioned earlier) and Ciao [13]. Ciao is a dataset collected from a website that contained users' ratings for items and users' comments and social connections.

In this paper, we randomly chose 75% of the users to be the training set and the rest were the test set. We set the size of the recommendation list from 3 to 10 and demonstrated the precision and recall rates as the number of recommended items increased.

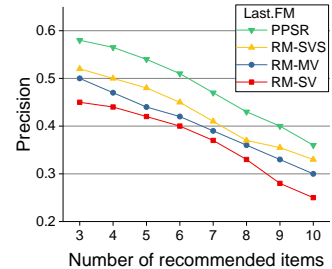
Fig. 2 Precision rates on Last.FM dataset

Fig. 2 shows the precision rates of our proposed PPSR scheme compared with three baseline methods on the Last.FM dataset, and Fig. 3 shows the precision rates of these four schemes on the Ciao dataset. It is apparent that our proposed PPSR scheme outperformed the baseline methods. In addition, methods with social data have better recommendation results than those without social information, and methods based on multi-view clustering obtain more latent information and have better performance than those based on single-view clustering. Since the Last.FM dataset consisted of more views of data than the Ciao dataset, the average recommendation precision of Last.FM was better than that of Ciao, which indicates that the density of a dataset leads to higher-quality recommendations.

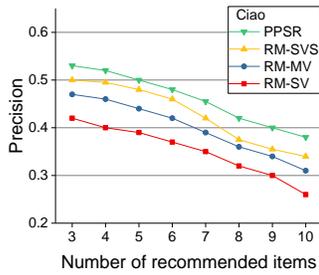

Fig. 3  Precision rates on Ciao dataset

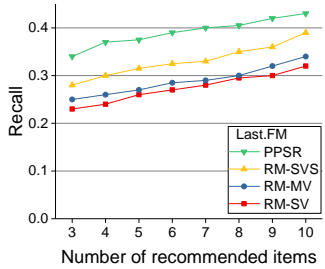

Fig. 4  Recall rates on Last.FM dataset

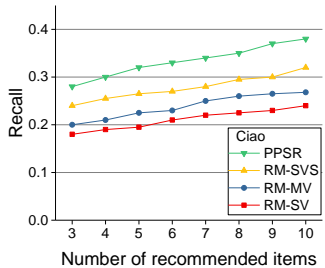

Fig. 5  Recall rates on Ciao dataset

Fig. 4 shows the recall rates of our proposed PPSR scheme compared with three baseline methods on Last.FM dataset, and Fig. 5 shows the precision rates of these four schemes on the Ciao dataset. Similar to the precision rates, our PPSR scheme also was the optimal scheme among these four models.

## V. CONCLUSION

In this paper, we introduced a Privacy-Preserving Socialized Recommendation (PPSR) model based on multi-view clustering to achieve secure and high-quality recommendations. SCMR combines the latent semantic information mined by the recommendation service provider and the social connections from the online social network provider to solve the Cold Start problem. Paillier homomorphic encryption further protects the privacy of social data and the similarity between users. The experimental results on different kinds of datasets demonstrated that PPSR is practical for real-world applications.

ACKNOWLEDGMENT

This paper is supported by the National Science Foundation of China under grant No. 61871064, 61501080, and 61771090, the Fundamental Research Funds for the Central Universities under No. DUT19JC08, and the China Postdoctoral Science Foundation under grant No. 2019M661097.